%% file: WZprd_final_eprint.tex
\documentclass[twocolumn,showpacs,preprintnumbers,amsmath,amssymb,superscriptaddress,showkeys]{revtex4}

\usepackage{graphicx}
\usepackage{dcolumn}
\usepackage{bm}

\newcommand{\pom}{I$\!\!$P}
\def\met{\not\!\!E_T}

\begin{document}

\hyphenation{PYTHIA DD SD ND}

\title{Diffractive $W$ and $Z$ Production at the Fermilab Tevatron}

\input{March2010_Authors}


\date{\today}

\pacs{14.70.Fm, 14.70.Hp, 12.40.Nn, 11.55.Jy}
\keywords{diffraction}


\begin{abstract}
We report on a measurement of the fraction of events with a $W$ or $Z$ boson which are produced diffractively in ${\bar p}p$ collisions at $\sqrt{s}=1.96$ TeV,  using data from 0.6 fb$^{-1}$ of integrated luminosity collected with the CDF~II detector equipped with a Roman-pot spectrometer that detects the $\bar p$ from $\bar{p}+p\rightarrow {\bar p}+[X+W/Z]$. We find that $(1.00\pm 0.11)\%$ of $W$s and $(0.88\pm 0.22)\%$ of $Z$s are produced diffractively in a region of antiproton {\em or proton} fractional momentum loss $\xi$ of $0.03<\xi<0.10$ and 4-momentum transferred squared $t$ of $-1<t<0$ (GeV/$c$)$^2$, where we account for the events in which the proton scatters diffractively while the antiproton dissociates, $\bar{p}+p\rightarrow [X+W/Z]+p$, by doubling the measured proton dissociation fraction. We also report on searches for $W$ and $Z$ production in double Pomeron exchange, $p+\bar{p}\rightarrow p+[X+W/Z]+\bar{p}$, and on exclusive $Z$ production, $\bar{p}+p\rightarrow {\bar p}+Z+p$. No signal is seen above background for these processes, and comparisons are made with expectations.
\end{abstract}

\maketitle
\section{Introduction}
Approximately one quarter of the inelastic $\bar{p}p$ collisions at the Tevatron are diffractive interactions, in which a strongly-interacting color singlet quark/gluon combination with the quantum numbers of the vacuum (the {\em Pomeron}, $\pom$) is presumed to be exchanged~\cite{Collins}-\cite{Donnachie}. As no radiation is expected from such an  exchange, a pseudorapidity region devoid of 
particles, called a {\em rapidity gap}~\cite{rapidity}, is produced. Diffractive processes are classified by the topology of the final state as single diffraction (SD), double Pomeron exchange (DPE), and double diffraction dissociation. 
In SD, the $\bar{p}(p)$ remains intact escaping the collision with momentum close to that of the original beam momentum and separated by a rapidity gap from the products of the $\pom$-$p(\bar{p})$ collision, usually referred to as a {\em forward~gap}; in DPE both the $\bar p$ and the $p$ escape, resulting in two forward rapidity gaps; and in double diffraction dissociation a central gap is formed while both the $p$ and $\bar p$ dissociate. 
A special case of rapidity gap events is {\em exclusive} production where a particle state is centrally produced, such as a dijet system or a $Z$ boson.

\begin{figure*}[ht] 
\centerline{\hspace*{2em}\includegraphics[angle=0,width=0.5\textwidth]{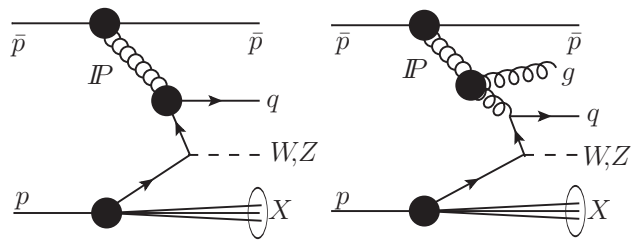}}
\caption{Diffractive $W$ production: (left) through quarks, and (right) through gluons.}
\label{fig:WZ_fig1} 
\end{figure*}

Diffraction has traditionally been described using 
Regge theory~(see Refs.~\cite{Collins}-\cite{Donnachie}).  In {\em hard} diffraction, such as jet or $W$ production, in addition to the colorless exchange there is a hard scale which allows one to 
explore both the mechanism for diffraction and the partonic structure of the Pomeron.
Whereas diffractive dijets can be produced via quarks or gluons, 
to leading order a diffractive $W$ is produced via a quark in the Pomeron.
Production by gluons is suppressed by a factor of $\alpha_s$, and can be distinguished from quark production by an additional 	associated jet, as shown in 
Fig.~\ref{fig:WZ_fig1}.
Combining cross section measurements of diffractive dijet production and diffractive $W$ production can be used to determine the quark/gluon content of the Pomeron~\cite{run1dijet1997}.
In Tevatron Run I, CDF measured the fraction of events with dijets~\cite{run1dijet1997}-\cite{run1dijet2002}, $W$ bosons~\cite{run1}, $b$ quarks~\cite{cdfdiffrb}, or $J/\psi$s~\cite{jpsi} which are produced diffractively, and found in all cases a fraction of approximately $1\%$ (including both $\pom$-$\bar p$ and $\pom$-$p$ production).

In the Run~I measurements, with the exception of ~\cite{run1dijet2000,run1dijet2002} which used a Roman-pot spectrometer (RPS), both the CDF and D0 collaborations used the rapidity gap signature for identifying diffractive events and extracting the diffractive fraction.
The interpretation of the results obtained by this method is complicated by the issue of gap survival probability, the likelihood that a rapidity gap produced in a diffractive interaction will not be filled by the products of additional parton-parton interactions in the same $\bar pp$ collision. 
In addition, there are experimental problems associated with the definition of the gap size, $\Delta\eta$, e.g., penetration of the gap by low transverse momentum ($p_T$) particles originating at the interaction point (IP) from the diffractively dissociated (anti)proton. To ameliorate this problem, CDF allowed up to two particles in the nominal gap region and introduced the term ``gap~acceptance'' for the fraction of events selected by this criterion. The gap survival and gap acceptance  probabilities both require model dependent Monte Carlo simulations followed by a detector simulation. 
At $\sqrt s=1.8$~TeV, CDF observed a diffractive $W$ signal with a probability $1.1\times 10^{-4}$ of being caused by a fluctuation from nondiffractive (ND) events and measured the fraction of diffractive $W$ events to be $[1.15\pm 0.51\mbox{(stat.)}\pm 0.20\mbox{(syst.)}]\%$~\cite{run1}. 
D0 studied both diffractive $W$ and $Z$ production in Run~I and found a diffractive fraction, uncorrected for gap survival, of $\left (0.89^{+0.19}_{-0.17}\right )\%$ for $W$s and $\left (1.44^{+0.61}_{-0.52}\right )\%$ for $Z$s~\cite{d0}. However, the gap survival estimated by D0 using Monte Carlo simulations was $(21\pm4)\%$, which would yield $W$ and $Z$ fractions approximately four times larger than those of CDF. An observation of an anomalously high diffractive $W/Z$ production rate could be evidence for beyond-standard-model theories, such as that of Ref.~\cite{AlanWhite} in which the Pomeron couples strongly to the electroweak sector through a pair of {\em sextet} quarks. The Run~I CDF and D0 measurements using rapidity gaps rely on model dependent corrections for the gap acceptance and background, making their interpretation difficult. The analysis presented here using the RPS makes no gap requirements and consequently is model-independent.  

\section{\label{sec:fwdde}Detector}
\begin{figure*} 
\centerline{\includegraphics[angle=0,width=0.8\textwidth]{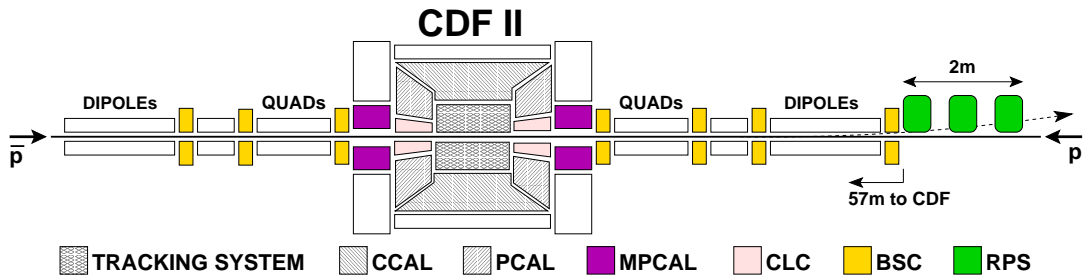}}
\caption{Plan view of the CDF~II detector (not to scale); the proton beam points to the $+\hat{z}$ (positive $\eta$) direction~\cite{rapidity}.}
\label{fig:WZ_fig2} 
\end{figure*}

The CDF II detector is a multi-purpose detector described in detail in Ref.~\cite{cdf}. It consists of a central detector and a forward detector system designed for diffractive physics studies (see Fig.~\ref{fig:WZ_fig2}).

The  central detector comprises a precision tracking system ($|\eta|\lesssim2$), central and plug calorimeters ($|\eta|<1.1 $ and $1.1<|\eta|<3.6$, respectively), with electromagnetic followed by hadronic sections, and muon spectrometers outside the central calorimeters ($|\eta|\lesssim 1.0$).  The tracking system is coaxial with the beam pipe and consists of silicon strip detectors surrounded by the Central Outer Tracker (COT), a cylindrical wire drift chamber inside a 1.4 T solenoidal magnetic field. 
 Proportional strip and wire chambers embedded inside the electromagnetic calorimeter provide an accurate position measurement of the source of electromagnetic showers.  Wire chambers and scintillator counters outside the calorimeters make up the muon detectors.

The forward detectors (see Ref.~\cite{DPEdijet}) consist of the Cherenkov Luminosity Counters (CLC) ($3.7<|\eta|<4.7$), which record charged particles coming from the IP and are used primarily for monitoring the luminosity; two MiniPlug calorimeters (MPCAL) covering the pseudorapidity region $3.5<|\eta|<5.1$; beam shower counters (BSC) within $5.4<|\eta|<7.4$ surrounding the beam pipe in several locations to detect charged particles and photons through conversion to $e^+e^-$ pairs in a 1 radiation length Pb plate placed in front of the first counter; the RPS, consisting of three Roman-pot (RP) detectors approximately 56~m from the nominal IP and 20~m from a string of Tevatron dipole magnets, used to detect and measure the momentum of diffracted antiprotons with fractional momentum loss $\xi$ in the region of $0.03\lesssim\xi\lesssim0.10$.  

\begin{figure} 
\centerline{\includegraphics[angle=0,width=0.4\textwidth]{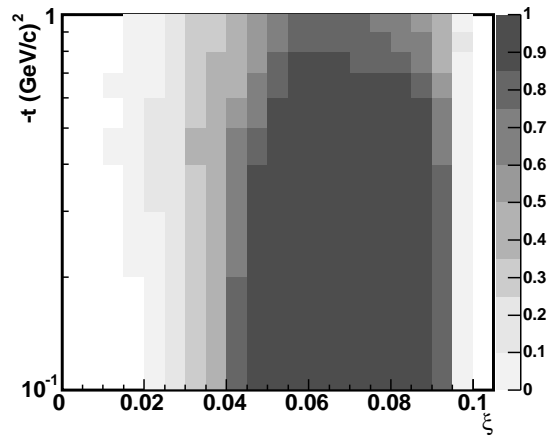}}
\caption{RPS acceptance as a function of $\xi$ and $t$ obtained from simulation 
using the transport parameters between the nominal interaction point and the Roman pots.}
\label{fig:WZ_fig3} 
\end{figure}

Each RPS detector consists of a trigger scintillation counter and clad scintillating fibers for tracking, covering 20 mm in  the $x$ direction to measure the distance from the beam in the plane of the Tevatron ring, and 20 mm in the $y$ direction perpendicular to the plane of the Tevatron. The fibers in both $x$ and $y$ in each RP were arranged in two layers, spaced by 1/3 of a fiber width to give three times better position resolution than a single layer~\cite{ken_thesis,run1dijet2000}. For RPS detector arrangement and track reconstruction details see Appendix C in Ref.~\cite{ken_thesis}. The resolutions in $\xi$ and $t$ (the 4-momentum transferred squared) for tracks recorded in the RPS were $\delta\xi=0.001$ and $\delta t=0.07$~(GeV/$c$)$^2$.
  The RPS acceptance is concentrated in the region of $0.03\lesssim\xi\lesssim 0.10$ and 
$-1<t<0$~(GeV/$c$)$^2$, as shown in Fig.~\ref{fig:WZ_fig3}.  
The RPS was installed for use in the Tevatron Run Ic (1995--96) and was operated in Run II from 2002 to February 2006.

A large fraction of events in which all three RPS trigger counters were hit are due to background which we refer to as {\em splash}\, events.  These  events are characterized by a large signal measured in the trigger counters as well as hits in almost all of the fibers.
One example of a splash process is a diffractive $\bar p$ which is outside 
the RPS acceptance showering in material near the RPS stations.  Another example is
beam halo induced interactions in nearby material.

The kinematics of the diffractive antiproton are reconstructed from the track position and angle in the RPS, and
the beam position and angle at the IP using a model of the Tevatron beam optics between the Roman pots and the IP. 

\section{\label{sec:datasets}$W$/$Z$ event selection}
The event selection begins by requiring a central ($|\eta|\!<\!1.1$) high-$p_T$ electron or muon consistent with
events with a $W$ or $Z$ decaying leptonically.
The triggering $e$($\mu$) is required to have a reconstructed $E^e_T(p^\mu_T)>20$ GeV~(GeV/$c$); for $Z$ candidates, the second lepton is subjected to  looser identification (ID) requirements, and electrons are also accepted if detected in the plug calorimeter within $1.2\!<|\eta|\!<2.8$.
Our data set with the forward detectors fully operational corresponds to an integrated luminosity of $\sim 0.6$ fb$^{-1}$ for both the electron and muon samples.

The $W$ candidate selection criteria require an $e$ or $\mu$ which passes tight ID requirements~\cite{Wcrosssect}
and has $E_T^e(p_T^\mu)>25$ GeV (GeV/$c$), 
missing transverse energy~\cite{missingET} (corrected for event vertex $z$-position, $p_T^\mu$ of muons that traverse the detector, and mismeasured hadronic jets) of $\met\,>25$ GeV, and $W$ transverse mass, $M_T^W=\sqrt{2(p_T^{l}p_T^{\nu}-\overrightarrow{p}_T^l\cdot\overrightarrow{p}_T^{\nu})}/c$,
in the region 
$40<M_{T}^{W}<120$~GeV/$c^2$.  The $Z\rightarrow ee$ selection criteria include the same 
requirements on the first electron plus a central $e$ with looser ID requirements, pertaining mainly to COT track quality and allowed range of ratio of energies deposited in the hadronic and electromagnetic calorimeters, or an $e$ in the plug calorimeter with $E_T^e>25$ GeV. Similarly, the $Z\rightarrow \mu\mu$ selection criteria include the same 
requirements on the first $\mu$ plus a loose-ID $\mu$ with $p_T^\mu>25$ GeV/$c$. For both the $Z\rightarrow ee$ and $Z\rightarrow \mu\mu$ channels, we also require $66<M_Z<116$~GeV/$c^2$.

The events are required to have a primary interaction vertex within 60 cm (in $z$) of the nominal IP.  Events with multiple vertices are not explicitly rejected in the event selection; instead, the fraction of events expected to have a single interaction is calculated.  The number of events passing the $W$ and $Z$ candidate selection requirements is 308~915 $W\rightarrow e\nu$, 259~465 $W\rightarrow \mu\nu$, 31~197 $Z\rightarrow ee$, and 15~603 $Z\rightarrow \mu\mu$.

The probability $P_n$ of a beam crossing producing $n$ inelastic interactions 
in addition to the hard interaction that produces the $W$/$Z$ is obtained from Poisson statistics: 
$P_n={\bar n}^n e^{-\bar n}/n!$, where ${\bar n}={\cal L}\cdot\sigma_{\rm inel}/f^{\rm cross}_{\rm eff}$ is the mean number
of interactions, which depends on the instantaneous luminosity ${\cal L}$, the inelastic cross section $\sigma_{\rm inel}$, and the effective beam crossing frequency (disregarding the transition regions of empty beam bunches) of $f^{\rm cross}_{\rm eff}=1.7$~MHz.
For the value of $\sigma_{\rm inel}$ at $\sqrt s=1.96$~TeV we use $59.3\pm 2.3$~mb, obtained by extrapolation from the CDF measurements of the elastic and total cross sections at $\sqrt s=1.8$~TeV~\cite{CDF_elastic,CDF_total}. The fraction of $W$ and $Z$ events from a single interaction
is determined independently for each one of three datasets comprising our event sample, collected  at different average instantaneous luminosities. This fraction is $f_{\rm 1-int}=(47.4\pm 1.3)\%$ for the Aug'02--Dec'03 dataset, $(25.1\pm 1.2)\%$ for Dec'04--Jul'05, and $(20.1\pm 1.0)\%$ for Sept'05--Feb'06, where the uncertainty is based on that in $\sigma_{\rm inel}$, which is common in all data sets.  The weighted average over all datasets is $\langle f_{\rm 1-int}\rangle =(25.6\pm 1.2)\%$. Taking into account bunch-to-bunch variation in luminosity, because our W/Z was more likely to be produced by a bunch with higher luminosity, we find that the single-interaction probability becomes systematically lower.  A correction factor of  $0.97\pm 0.01$ is applied to $\langle f_{\rm 1-int}\rangle$ to account for this effect.

\section{\label{sec:dataanal}Diffractive event selection} 
Diffractive events are first selected by requiring that all three of the RPS trigger 
counters have energy deposited within a specified run-dependent range.  Although a minimum
energy is required in order to select events where the RPS detectors are hit, 
we also require a reconstructed RPS track.  
The upper bound on the energy imposed in order to remove background splash events is another important selection requirement.
As splash events are caused by 
secondaries from an interaction in material near the RPS,
rather than by a diffractive antiproton within the RPS acceptance, these events tend to have
large energy deposited in the RPS trigger counters, as well as a large fraction of RPS 
tracking fibers hit.
Next, a reconstructed RPS track is also required.  
In this step, we accept events where the $\xi$ and $t$ reconstructed from the track are within the range of $0.03<\xi<0.10$ and $|t|<1$~(GeV/$c$)$^2$.  

In an event with multiple ${\bar p}p$ interactions, the diffracted antiproton may originate from a different interaction than the one producing the $W$ or $Z$. This {\em overlap background} is dominated by events in which a ND $W/Z$ is superimposed on an inclusive SD interaction with the $\bar p$ detected in the RPS. The main tool for removing overlap backgrounds is the value of $\xi_{\bar p}$ reconstructed using the calorimeters,  

\begin{equation}\label{eq:xical}\xi^{\rm cal}_{\bar p}=\sum_{i=1}^{N_{\rm towers}}\frac{E^i_{\rm T}}{\sqrt{s}}e^{-\eta^i},
\end{equation}

\noindent where $\eta^i$ is the $\eta$--value of the center of a tower and the sum is carried over all calorimeter towers with $E_T$ greater than a calorimeter-dependent threshold chosen to reject noise (see Ref.~\cite{DPEdijet}). 

The $\xi^{\rm cal}_{\bar p}$ values were calibrated by comparing diffractive dijet data, collected concurrently with the diffractive $W/Z$ data, with Monte Carlo generated events.  Calibrated $\xi^{\rm cal}_{\bar p}$ values were found to be in good agreement with values measured by the RPS.  
The resolution in $\xi^{\rm cal}_{\bar p}$ is dominated by that of the energy measurement by MPCAL, which is $\sim 30$\% resulting in a root mean square deviation $\Delta\log\xi=0.1$.
In a diffractive Z event with no additional interactions, $\xi^{\rm cal}_{\bar p}\approx \xi^{\rm RPS}$ and should fall within the RPS acceptance region of $0.03<\xi<0.10$ or $-1.5<\log{\xi}<-1.0$.  In an event with multiple interactions, $\xi^{\rm cal}_{\bar p}$ would be larger due to energy from the additional interaction.  Therefore, we expect all diffractive Z interactions with no additional interaction to have $\xi^{\rm cal}_{\bar p}<0.10$ and we can remove most of the overlap background with this requirement.  Some events with $\xi^{\rm cal}_{\bar p}<0.10$ may still have multiple interactions.  We use the distribution of $\xi^{\rm cal}_{\bar p}$ from the nondiffractive Z sample (before the requirement of a RPS track) to estimate the overlap background at small $\xi^{\rm cal}_{\bar p}$.  One does not expect to find perfect agreement between the ND and SD distributions at $\xi^{\rm cal}_{\bar p}>0.1$ since the SD candidates in that region always contain at least one other interaction, while a fraction of the ND events may be due to a single interaction. By normalizing the ND to the SD distribution in the region of $-1.0<\log{\xi}<-0.4$ we obtain a reasonable estimate of the overlap background in the region $\xi^{cal}_{\bar p}<0.1$ within the assigned conservative uncertainty.

\begin{figure*}
\centerline{\includegraphics[angle=0,width=0.7\textwidth]{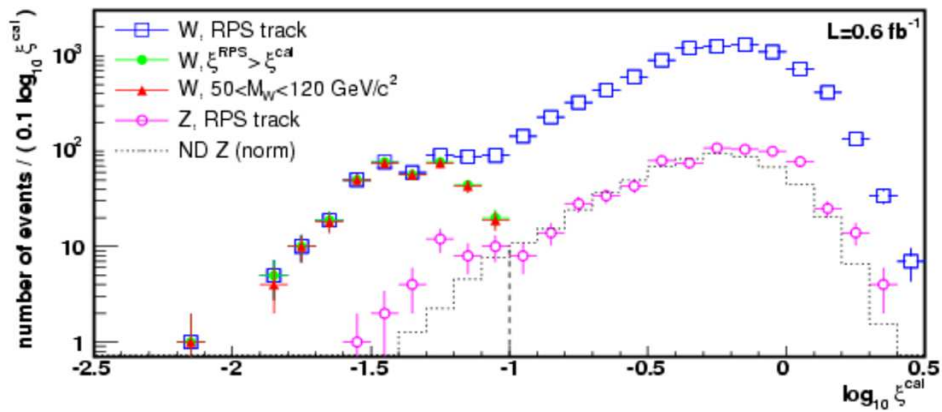}}
\caption{$\xi_{\bar p}^{\rm cal}$ for $W$ and $Z$ events with a RPS track. The dotted histogram is the distribution of ND $Z$ events normalized to the data $Z$-distribution in the region $-1.0<\log_{10}\xi^{\rm cal}_{\bar p}<-0.4$.}
\label{fig:WZ_fig4} 
\end{figure*}

Figure~\ref{fig:WZ_fig4} shows the distributions of $\xi^{\rm cal}_{\bar p}$ for $W$ and $Z$ candidate events with a RPS track. For $Z$s, the distribution for ND events is also shown, normalized to the distribution from events with a RPS track in the region of $-1.0<\log_{10}\xi^{\rm cal}_{\bar p}<-0.4$. The excess of RPS events over ND ones for $\log_{10}\xi^{\rm cal}_{\bar p}<-1.0$ ($\xi_{\bar p}^{cal}<0.10$) contains the SD signal. In determining the fraction of $Z$ events which are diffractive in Sec.~\ref{WZfraction}, we require $\xi^{cal}_{\bar p}<0.10$ (number of events $N^Z_{\xi<0.10}$) and subtract a background determined from the number of events expected in the normalized ND distribution with $\xi^{cal}_{\bar p}<0.10$ ($N^Z_{bgnd}$). 

In diffractive $W$ events, since the neutrino is not directly detected by the CDF~II detector, $E_T^\nu$ is not included in the $\sum E_T$
of calorimeter towers used to determine $\xi_{\bar p}^{\rm cal}$ through Eq.~(\ref{eq:xical}).  
Requiring $\xi_{\bar p}^{\rm cal}<\xi_{\bar p}^{\rm RPS}$ removes  events with multiple interactions in a similar
way as the requirement on $\xi_{\bar p}^{cal}<0.10$.
In addition, knowing the kinematics of the
diffracted antiproton allows us to reconstruct the longitudinal momentum of the neutrino and thereby the $W$
mass, $M_{\rm W}$, as described in Sec.~\ref{sec:Wkin}.  
Requiring $M_{\rm W}$ to be
within the range $50<M_{\rm W}<120$ GeV/$c^2$ removes almost all the remaining multiple-interaction or misreconstructed events.

The $\xi_{\bar p}^{\rm cal}$ distribution for diffractive $W$ candidates is shown in Fig.~\ref{fig:WZ_fig4}, and the number of $W$ and $Z$ events passing the diffractive event selection criteria is 
listed in Table~\ref{t:diffrWZ}.

\begin{table}[ht]
\begin{center} 
\caption{\label{t:diffrWZ} $W$ and $Z$ events passing successive selection requirements.}
\vspace*{0.5em}
\begin{tabular}{lrrr}\hline
\hline

 & {$W\rightarrow e\nu$} & {$W\rightarrow \mu\nu$} & $W\rightarrow l(e/\mu)\nu$ \\

RPS-trigger-counters & 6663 & 5657 & 12~320 \\
RPS-track     & 5124 & 4201 &  9325 \\
$50<M_W<120$ &  192  &  160 &   352 \\

 & {$Z\rightarrow ee$} & {$Z\rightarrow \mu\mu$} & $Z\rightarrow ll$ \\

RPS-trigger-counters & 650 & 341 & 991 \\
RPS-track   & 494 & 253 & 747 \\
$\xi^{\rm cal}<0.10$ &  24 &  12 &  36 \\
\hline
\hline
\end{tabular}
\end{center}
\end{table}

\section{Results}
\subsection{$W$ mass from diffractive events\label{sec:Wkin}}
In nondiffractive $W$ production, the neutrino transverse energy $E^\nu_T$ is inferred from the $\met$ but the neutrino longitudinal momentum $p^\nu_z$ is unknown. However, in diffractive $W$ production the missing $p^\nu_z$ yields a 
difference between the $\xi_{\bar p}^{\rm cal}$, calculated from the energy deposited in the calorimeters using Eq.~(\ref{eq:xical}), and the $\xi^{\rm RPS}_{\bar p}$ determined from the RPS track.
This difference allows one to determine  $p^\nu_z$, and thereby the full $W$ kinematics through Eqs.~(\ref{eq:etanu}--\ref{eq:pEw}): 

\begin{equation}
\xi^{\rm RPS}-\xi_{\bar p}^{\rm cal}=\sum_{i=1}^{\rm all\;towers}\frac{\met^i}{\sqrt{s}}e^{-\eta^\nu}\label{eq:etanu},
\end{equation}

\begin{equation}
p_z^\nu=\met/\tan{\left[2\tan^{-1}{(e^{-\eta^\nu})}\right]}\label{eq:pznu},
\end{equation}

\begin{equation}
M_{\rm w}^2=2\left [E_e\sqrt{\met^2+p_z^{\nu\,2}}-p_x^e p_x^\nu-p_y^e p_y^\nu-p_z^e p_z^\nu\right]\label{eq:mw},
\end{equation}

\begin{equation}
p_z^{\rm w}=p_z^e+p_z^\nu,\; E^{\rm w}=E^e+\sqrt{\met^2+p_z^{\nu\,2}}.\label{eq:pEw}
\end{equation}

\begin{figure}[!htb]
\centerline{\includegraphics[angle=0,width=0.45\textwidth]{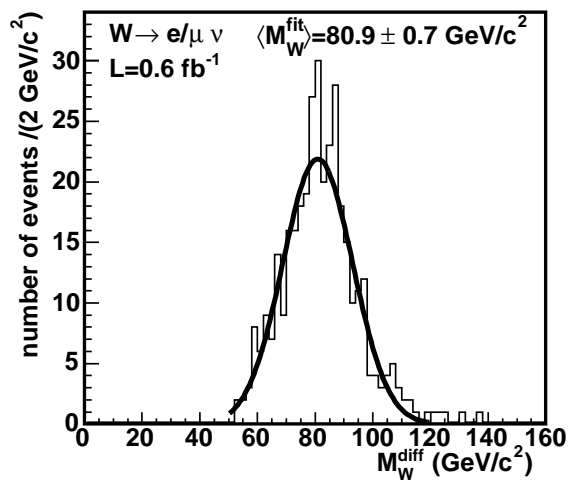}}
\caption{Reconstructed $M_{\rm W}^{\rm diff}$ with a Gaussian fit.}
\label{fig:WZ_fig5}
\end{figure}

Using the full $W$ kinematics results in a Gaussian $M_{\rm W}$ distribution permitting a more accurate determination of $M_{\rm W}$ from a given number of events.  
For the diffractive sample of 352 events listed in Table~\ref{t:diffrWZ}, this method yields the mass distribution shown in Fig.~\ref{fig:WZ_fig5} from which we obtain $M_{\rm W}^{\rm diff}=(80.9\pm0.7)$~GeV/$c^2$, in agreement with the world average of 
$M_W^\text{PDG}=(80.398\pm0.025)$~GeV/$c^2$~\cite{PDG}. 

\subsection{\label{WZfraction}Diffractive $W/Z$ fraction}

The RPS acceptance is calculated using:

\begin{equation}
A_{\rm RPS}=\frac{N}{\sum_{i=1}^{\rm N} A^i_{\rm RPS}(\xi_i,t_i)^{-1}},
\end{equation}

\noindent where the sum is over $N$ diffractive $W$ events with a RPS track and $A^i_{\rm RPS}(\xi,t)$ is the acceptance for an event at ($\xi_i,t_i$) shown in Fig.~\ref{fig:WZ_fig3}. 
For our diffractive $W$ event samples, we measure 
$A_{\rm RPS}=[88\pm 12(\rm stat.)]\%$ (Aug'02--Dec'03) and  $A_{\rm RPS}=[75\pm 5(\rm stat.)]\%$ (Dec'04--Feb'06) for events within $0.03<\xi<0.10$ and $|t|<1$ (GeV/$c$)$^{2}$. 
These values are consistent with those determined with better statistical precision from the data of our exclusive dijet production paper~\cite{DPEdijet}, which are mainly due to variations in beam angle dispersion at the collision point and RPS alignment over the period of data taking.
As the dijet data were taken concurrently with the diffractive $W$/$Z$ data, we use the acceptances of the dijet paper with an increased systematic uncertainty to account for differences in the $\xi$ and $t$ distributions expected between $W$/$Z$ and dijet production. The values being used are:

\begin{equation}
A_{\rm RPS}= [83\pm 5(\rm syst.)]\%\mbox{~(Aug'02--Dec'03),}
\end{equation}
\begin{equation}  
A_{\rm RPS}= [78\pm 5(\rm syst.)]\%\mbox{~(Dec'04--Feb'06).}
\end{equation}

The requirement on RPS trigger counter energy was made to remove splash events due to beam-related background and events with diffractive antiprotons outside the RPS acceptance.  However, it also removes some diffractive signal events.  
The efficiency for this selection requirement to retain good diffractive events is determined as the fraction of all events with a reconstructed RPS track which pass the requirement on trigger counter energy.  This efficiency, $\epsilon_{\rm RPStrig}$, varies with dataset within the range $68\%-80\%$, as the selection requirements were chosen independently for different running periods 
to account for ageing of the counters.  
The average RPS track reconstruction efficiency is~\cite{DPEdijet} 
 
\begin{equation}
\epsilon_{\rm RPStrk}=[87\pm 1(\mbox{stat.})\pm 6(\mbox{syst.})]\%,
\end{equation}

\noindent where the systematic uncertainty was chosen to bring consistency over all data sets.

\subsubsection{Double Pomeron exchange\label{DPE/SD}}
Double Pomeron exchange events are the sub-sample of the $W/Z$ SD events  in which both the $\bar p$ and $p$ remain intact. 
As for the SD events, defined by the requirement of $\xi_{\bar p}^{\rm cal}\leqslant 0.1$, a DPE interaction should have in addition $\xi^{\rm cal}_{p}\leqslant 0.1$, where
$\xi^{\rm cal}_{p}=\sum_{i=1}^{\rm all-towers}(E^i_T/ \sqrt{s})\,e^{+\eta^i}$.
In this region of $\xi^{\rm cal}_{p}$ there are 45 $W$ and two $Z$ events in our candidate $W/Z$ data. 
Monte Carlo studies show that these are consistent with the expected numbers of SD events in which the gap on the proton side is due to multiplicity fluctuations, without any DPE contribution. Using fits of Monte Carlo templates to the data conducted according to Ref.~\cite{Barlow}, profile likelihood limits were set at the 95\% confidence level of 1.5\% and 7.8\% on the fraction of diffractive $W$ and $Z$ events produced by DPE, respectively. 
These limits are consistent with the expectation of no observable signal due to the reduced collision s-value from 0.1s ($\sqrt s\sim 600$~GeV) in SD to $(0.1)^2s$ ($\sqrt s\sim 200$~GeV) in DPE.
 
\subsubsection{Search for exclusive $Z$ production\label{exclZsearch}}
We have examined whether any of the $Z$ candidate events in the DPE event sample are produced exclusively through the process ${\bar p}+p\rightarrow {\bar p}+Z+p$.  In the Standard Model, this process is predicted to proceed by photoproduction, where a virtual photon radiated from the $\bar p$ ($p$) fluctuates into a $q\bar q$ loop which scatters elastically on the $p$ ($\bar p$) by two-gluon exchange forming a $Z$. Since $W$s cannot be produced exclusively, because there must be at least another charged particle in the event, we use $W$s as a control sample.

A limit on exclusive $Z$ production has recently been published by the CDF collaboration. At a 95\% confidence level, the exclusive $Z$ production cross section was found to be  $\sigma_{\rm excl}^Z<0.96$~pb, a factor of ${\cal{O}}(10^{-3})$ of predictions based on the Standard Model (see Ref. ~\cite{exclZ}-\cite{albrow_forshaw}). The search method relied on strict excusivity requirements to ensure that nothing is present in the detector except for the two leptons from $Z\rightarrow l^{+l}l^{-}$.

The present search is based on our data sample of $0.6$ fb$^{-1}$ integrated luminosity, which is a $\sim 30\%$ sub-sample of the events used in the above search. However, we use a different method which is more transparent to the background introduced by detector noise that can spoil the exclusivity requirement, and expand our event selection beyond the $Z$-mass window to include all $ee$($\mu\mu$) pairs with an electron~(muon) $E_T(p_T)>25$ GeV(GeV/$c$).
The method consists of comparing the total energy in the calorimeter to the dilepton mass. At first, we determine the dilepton mass $M_{ll}$ from the $e$ or $\mu$ momentum using calorimeter ($e$) or track ($\mu$) information.  For $W$s, we evaluate $M_W$ using Eq.~(\ref{eq:mw}).  Then, we determine the system mass $M_X$ from 
calorimeter towers, $M_X=\sqrt{(\Sigma E)^2-(\Sigma {\vec p})^2}$, with the caveat that we substitute the muon track information for the
corresponding calorimeter tower, and in the case of $W$ events include the $\nu$ momentum.
Thus, if nothing is present in the calorimeter besides the dilepton pair (or the $W$), the event will satisfy the condition $M_X=M_{ll}$ (or $M_X=M_W$).

\begin{figure}[!htb]
\begin{center}
\centerline{\hspace*{2.6em}\includegraphics[angle=0,width=0.28\textwidth]{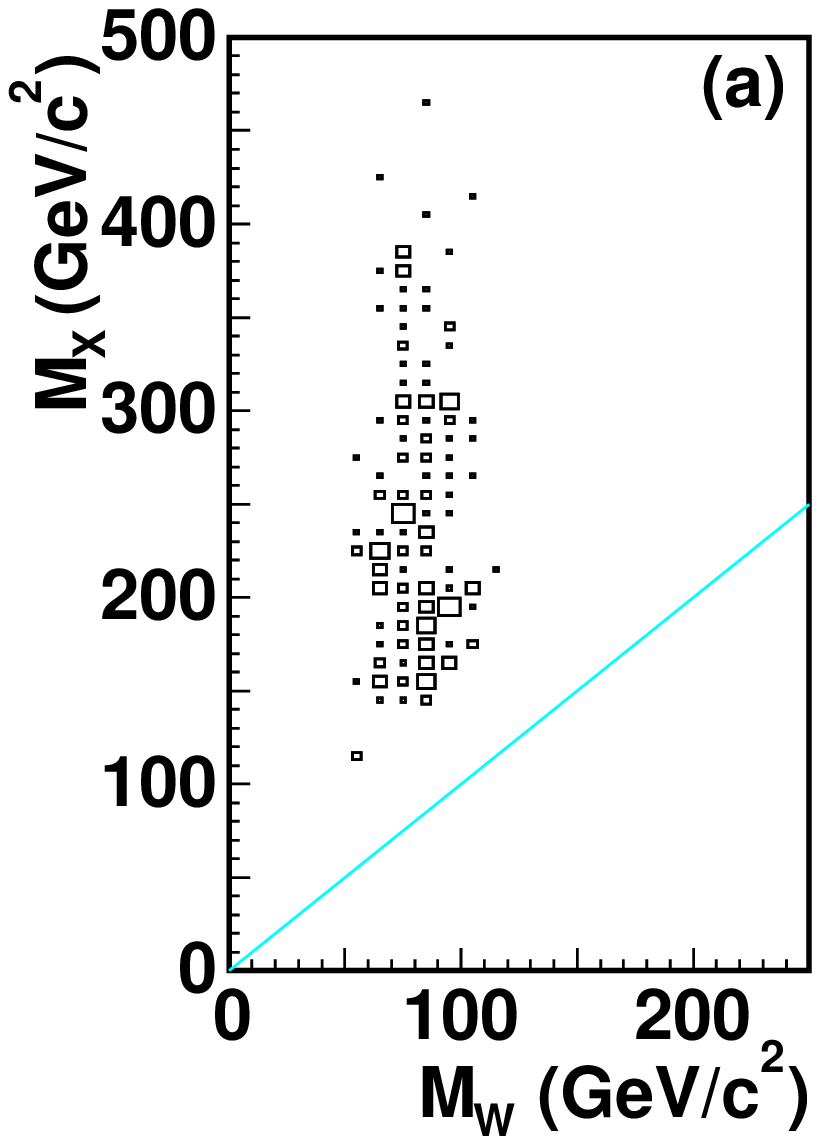}\hspace*{-1em}\includegraphics[angle=0,width=0.28\textwidth]{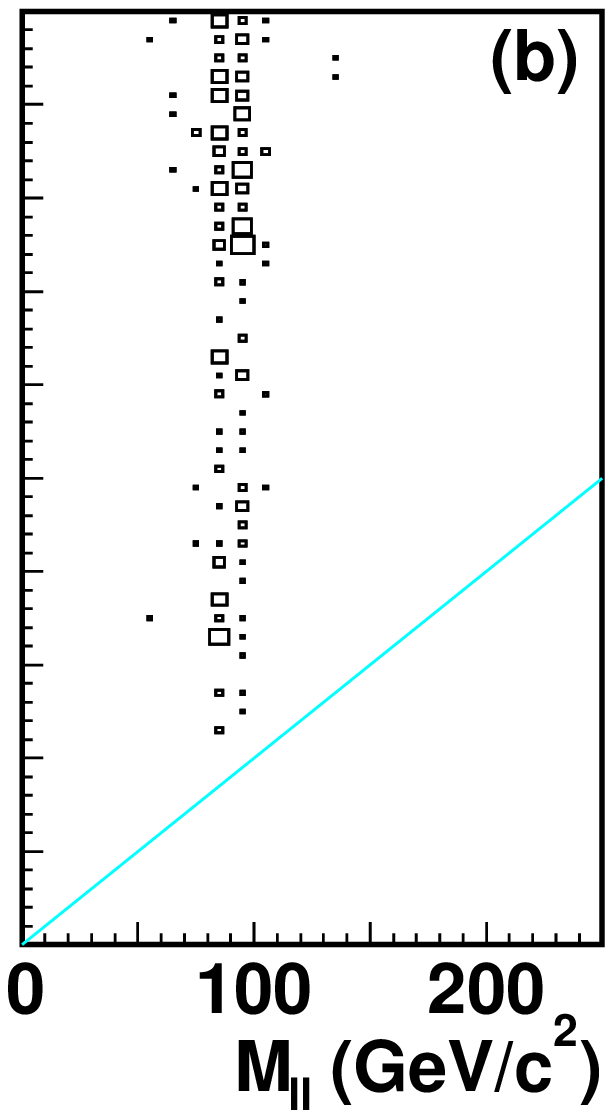}}
\caption{System mass $M_X$ vs $M_W$ (left) for $W$ events with a RPS track and $\xi^{\rm cal}<\xi^{\rm RPS}$, and mass $M_{ll}$ (right) for dilepton events with a RPS track and lepton $p_T>25$ GeV/$c$. Exclusive $W/Z$ candidates are expected to fall on the $M_X=M_W\,(M_{ll})$ line.}
\label{fig:WZ_fig6}
\end{center}
\vspace*{-1cm}\end{figure}

Figure~\ref{fig:WZ_fig6} shows $M_X$ versus $M_W$ ($M_{ll}$) for $W$ (dilepton) events with a RPS track.  Note that $M_X$ depends on the calorimeter thresholds: increasing the thresholds could cause more events to move onto the $M_X=M_W\,(M_{ll})$ line, but these events would not be truly exclusive.  
The diffractive $W$ sample acts as a control sample for this type of background.  Calorimeter noise could also move truly exclusive events off the $M_X=M_W\,(M_{ll})$ line.  
We looked at ``empty crossings" collected using a beam-crossing trigger and selecting events with no reconstructed tracks or hits in the Cherenkov luminosity counters.  The energy and momentum of the ``noise" in the calorimeter for these events was added to the candidate exclusive events and the difference in reconstructed $M_X$ determined.  
The largest mean deviation found in the different running periods was 0.5 GeV, which is represented by the width of the $M_X=M_W\,(M_{ll})$ line in Fig.~\ref{fig:WZ_fig6}.
No candidates for exclusive production are observed within this band, neither in the control sample of the $W$ events (left)  nor in the $Z$ event sample (right). This result is compatible with the upper bound on exclusive $Z$ production set in Ref.~\cite{exclZ}, as expected.        

\subsubsection{Diffractive fractions} The diffractive fractions $R_{\rm W}$ and $R_{\rm Z}$ for $0.03\!<\!\xi\!<\!0.10$ and $|t|\!<\!1$ (GeV/$c$)$^2$ are obtained after RPS acceptance and background corrections. 
The corrections include division by the product of the RPS acceptance, $A_{\rm RPS}$, the efficiency of the selection requirement on RPS trigger counter energy, $\epsilon_{\rm RPStrig}$, the RPS tracking efficiency, $\epsilon_{\rm RPStrk}$, and the fraction of ND  events which are expected to have a single interaction, $N^{\rm 1-int}_{ND}=N_{ND}\cdot \langle f_{\rm 1-int}\rangle$. To account for SD events in which the proton, instead of the antiproton, remains intact we multiply $N^W_{SD}$ or $N^Z_{SD}\equiv N^Z_{\xi<0.10}-N^Z_{\rm bgnd}$ by a factor of two.  

\begin{equation}
R_W(R_Z)=\frac
{
2\cdot N_{SD}^W(N_{SD}^Z)
}
{
A_{\rm RPS}\cdot \epsilon_{\rm RPStrig}\cdot \epsilon_{\rm RPStrk} \cdot N^{\rm 1-int}_{ND}
}
\end{equation}

The resulting diffractive fractions are:

\begin{eqnarray*}
R_{\rm W}=[1.00 \pm 0.05\,(\mbox{stat.}) \pm 0.10(\,\mbox{syst.})]\%,\\
R_{\rm Z}=[0.88 \pm 0.21(\,\mbox{stat.}) \pm 0.08(\,\mbox{syst.})]\%,
\end{eqnarray*}

\noindent where the systematic uncertainties are obtained from the contributions of the RPS acceptance and the trigger and tracking efficiencies.

\vspace*{-0.25cm}\section{Conclusions}
We have measured the fraction of events with a $W$ or $Z$ boson which are produced diffractively in ${\bar p}p$ collisions at $\sqrt{s}=1.96$ TeV using data from 0.6 fb$^{-1}$ of integrated luminosity collected with the CDF~II detector incorporating a Roman-pot spectrometer to detect diffracted antiprotons. Within a region of antiproton {\em or proton} fractional momentum loss $\xi$ of $0.03<\xi<0.10$ and 4-momentum transferred squared $t$ of $-1<t<0$ (GeV/$c$)$^2$, we find that $(1.00\pm 0.11)\%$ of $W$s and $(0.88\pm 0.22)\%$ of $Z$s are produced diffractively, where the events in which the proton scatters diffractively and the antiproton dissociates are accounted for by doubling the measured proton dissociation fraction.
We have also conducted a search for $W$ and $Z$ events produced by double-Pomeron exchange, $\bar p+p\rightarrow \bar p+[X+W/Z]+p$, and set confidence level upper limits of 1.5\% and 7.7\% on the fraction of DPE/SD events, respectively.
Finally, we searched for exclusive $Z$ production, $\bar p+p\rightarrow \bar p+Z+p$. No exclusive $Z$ candidates were found within the DPE event sample, compatible with the CDF Run~II published limit. 

\section{Acknowledgments}
We thank the Fermilab staff and the technical staffs of the participating institutions for their vital contributions. This work was supported by the U.S. Department of Energy and National Science Foundation; the Italian Istituto Nazionale di Fisica Nucleare; the Ministry of Education, Culture, Sports, Science and Technology of Japan; the Natural Sciences and Engineering Research Council of Canada; the National Science Council of the Republic of China; the Swiss National Science Foundation; the A.P. Sloan Foundation; the Bundesministerium f\"ur Bildung und Forschung, Germany; the World Class University Program, the National Research Foundation of Korea; the Science and Technology Facilities Council and the Royal Society, UK; the Institut National de Physique Nucleaire et Physique des Particules/CNRS; the Russian Foundation for Basic Research; the Ministerio de Ciencia e Innovaci\'{o}n, and Programa Consolider-Ingenio 2010, Spain; the Slovak R\&D Agency; and the Academy of Finland.

\end{document}

%% file: March2010_Authors.tex
\affiliation{Institute of Physics, Academia Sinica, Taipei, Taiwan 11529, Republic of China} 
\affiliation{Argonne National Laboratory, Argonne, Illinois 60439, USA} 
\affiliation{University of Athens, 157 71 Athens, Greece} 
\affiliation{Institut de Fisica d'Altes Energies, Universitat Autonoma de Barcelona, E-08193, Bellaterra (Barcelona), Spain} 
\affiliation{Baylor University, Waco, Texas 76798, USA} 
\affiliation{Istituto Nazionale di Fisica Nucleare Bologna, $^{bb}$University of Bologna, I-40127 Bologna, Italy} 
\affiliation{Brandeis University, Waltham, Massachusetts 02254, USA} 
\affiliation{University of California, Davis, Davis, California 95616, USA} 
\affiliation{University of California, Los Angeles, Los Angeles, California 90024, USA} 
\affiliation{Instituto de Fisica de Cantabria, CSIC-University of Cantabria, 39005 Santander, Spain} 
\affiliation{Carnegie Mellon University, Pittsburgh, Pennsylvania 15213, USA} 
\affiliation{Enrico Fermi Institute, University of Chicago, Chicago, Illinois 60637, USA}
\affiliation{Comenius University, 842 48 Bratislava, Slovakia; Institute of Experimental Physics, 040 01 Kosice, Slovakia} 
\affiliation{Joint Institute for Nuclear Research, RU-141980 Dubna, Russia} 
\affiliation{Duke University, Durham, North Carolina 27708, USA} 
\affiliation{Fermi National Accelerator Laboratory, Batavia, Illinois 60510, USA} 
\affiliation{University of Florida, Gainesville, Florida 32611, USA} 
\affiliation{Laboratori Nazionali di Frascati, Istituto Nazionale di Fisica Nucleare, I-00044 Frascati, Italy} 
\affiliation{University of Geneva, CH-1211 Geneva 4, Switzerland} 
\affiliation{Glasgow University, Glasgow G12 8QQ, United Kingdom} 
\affiliation{Harvard University, Cambridge, Massachusetts 02138, USA} 
\affiliation{Division of High Energy Physics, Department of Physics, University of Helsinki and Helsinki Institute of Physics, FIN-00014, Helsinki, Finland} 
\affiliation{University of Illinois, Urbana, Illinois 61801, USA} 
\affiliation{The Johns Hopkins University, Baltimore, Maryland 21218, USA} 
\affiliation{Institut f\"{u}r Experimentelle Kernphysik, Karlsruhe Institute of Technology, D-76131 Karlsruhe, Germany} 
\affiliation{Center for High Energy Physics: Kyungpook National University, Daegu 702-701, Korea; Seoul National University, Seoul 151-742, Korea; Sungkyunkwan University, Suwon 440-746, Korea; Korea Institute of Science and Technology Information, Daejeon 305-806, Korea; Chonnam National University, Gwangju 500-757, Korea; Chonbuk National University, Jeonju 561-756, Korea} 
\affiliation{Ernest Orlando Lawrence Berkeley National Laboratory, Berkeley, California 94720, USA} 
\affiliation{University of Liverpool, Liverpool L69 7ZE, United Kingdom} 
\affiliation{University College London, London WC1E 6BT, United Kingdom} 
\affiliation{Centro de Investigaciones Energeticas Medioambientales y Tecnologicas, E-28040 Madrid, Spain} 
\affiliation{Massachusetts Institute of Technology, Cambridge, Massachusetts 02139, USA} 
\affiliation{Institute of Particle Physics: McGill University, Montr\'{e}al, Qu\'{e}bec, Canada H3A~2T8; Simon Fraser University, Burnaby, British Columbia, Canada V5A~1S6; University of Toronto, Toronto, Ontario, Canada M5S~1A7; and TRIUMF, Vancouver, British Columbia, Canada V6T~2A3} 
\affiliation{University of Michigan, Ann Arbor, Michigan 48109, USA} 
\affiliation{Michigan State University, East Lansing, Michigan 48824, USA}
\affiliation{Institution for Theoretical and Experimental Physics, ITEP, Moscow 117259, Russia}
\affiliation{University of New Mexico, Albuquerque, New Mexico 87131, USA} 
\affiliation{Northwestern University, Evanston, Illinois 60208, USA} 
\affiliation{The Ohio State University, Columbus, Ohio 43210, USA} 
\affiliation{Okayama University, Okayama 700-8530, Japan} 
\affiliation{Osaka City University, Osaka 588, Japan} 
\affiliation{University of Oxford, Oxford OX1 3RH, United Kingdom} 
\affiliation{Istituto Nazionale di Fisica Nucleare, Sezione di Padova-Trento, $^{cc}$University of Padova, I-35131 Padova, Italy} 
\affiliation{LPNHE, Universite Pierre et Marie Curie/IN2P3-CNRS, UMR7585, Paris, F-75252 France} 
\affiliation{University of Pennsylvania, Philadelphia, Pennsylvania 19104, USA}
\affiliation{Istituto Nazionale di Fisica Nucleare Pisa, $^{dd}$University of Pisa, $^{ee}$University of Siena and $^{ff}$Scuola Normale Superiore, I-56127 Pisa, Italy} 
\affiliation{University of Pittsburgh, Pittsburgh, Pennsylvania 15260, USA} 
\affiliation{Purdue University, West Lafayette, Indiana 47907, USA} 
\affiliation{University of Rochester, Rochester, New York 14627, USA} 
\affiliation{The Rockefeller University, New York, New York 10065, USA} 
\affiliation{Istituto Nazionale di Fisica Nucleare, Sezione di Roma 1, $^{gg}$Sapienza Universit\`{a} di Roma, I-00185 Roma, Italy} 

\affiliation{Rutgers University, Piscataway, New Jersey 08855, USA} 
\affiliation{Texas A\&M University, College Station, Texas 77843, USA} 
\affiliation{Istituto Nazionale di Fisica Nucleare Trieste/Udine, I-34100 Trieste, $^{hh}$University of Trieste/Udine, I-33100 Udine, Italy} 
\affiliation{University of Tsukuba, Tsukuba, Ibaraki 305, Japan} 
\affiliation{Tufts University, Medford, Massachusetts 02155, USA} 
\affiliation{Waseda University, Tokyo 169, Japan} 
\affiliation{Wayne State University, Detroit, Michigan 48201, USA} 
\affiliation{University of Wisconsin, Madison, Wisconsin 53706, USA} 
\affiliation{Yale University, New Haven, Connecticut 06520, USA} 
\author{T.~Aaltonen}
\affiliation{Division of High Energy Physics, Department of Physics, University of Helsinki and Helsinki Institute of Physics, FIN-00014, Helsinki, Finland}
\author{B.~\'{A}lvarez~Gonz\'{a}lez$^v$}
\affiliation{Instituto de Fisica de Cantabria, CSIC-University of Cantabria, 39005 Santander, Spain}
\author{S.~Amerio}
\affiliation{Istituto Nazionale di Fisica Nucleare, Sezione di Padova-Trento, $^{cc}$University of Padova, I-35131 Padova, Italy} 

\author{D.~Amidei}
\affiliation{University of Michigan, Ann Arbor, Michigan 48109, USA}
\author{A.~Anastassov}
\affiliation{Northwestern University, Evanston, Illinois 60208, USA}
\author{A.~Annovi}
\affiliation{Laboratori Nazionali di Frascati, Istituto Nazionale di Fisica Nucleare, I-00044 Frascati, Italy}
\author{J.~Antos}
\affiliation{Comenius University, 842 48 Bratislava, Slovakia; Institute of Experimental Physics, 040 01 Kosice, Slovakia}
\author{M.G.~Albrow}
\affiliation{Fermi National Accelerator Laboratory, Batavia,  Illinois 60510, USA}
\author{G.~Apollinari}
\affiliation{Fermi National Accelerator Laboratory, Batavia, Illinois 60510, USA}
\author{J.A.~Appel}
\affiliation{Fermi National Accelerator Laboratory, Batavia, Illinois 60510, USA}
% ...
\collaboration{CDF Collaboration\footnote{With visitors from $^a$University of Massachusetts Amherst, Amherst, Massachusetts 01003,
$^b$Istituto Nazionale di Fisica Nucleare, Sezione di Cagliari, 09042 Monserrato (Cagliari), Italy,
$^c$University of California Irvine, Irvine, CA  92697, 
$^d$University of California Santa Barbara, Santa Barbara, CA 93106
$^e$University of California Santa Cruz, Santa Cruz, CA  95064,
$^f$CERN,CH-1211 Geneva, Switzerland,
$^g$Cornell University, Ithaca, NY  14853, 
$^h$University of Cyprus, Nicosia CY-1678, Cyprus, 
$^i$University College Dublin, Dublin 4, Ireland,
$^j$University of Fukui, Fukui City, Fukui Prefecture, Japan 910-0017,
$^k$Universidad Iberoamericana, Mexico D.F., Mexico,
$^l$Iowa State University, Ames, IA  50011,
$^m$University of Iowa, Iowa City, IA  52242,
$^n$Kinki University, Higashi-Osaka City, Japan 577-8502,
$^o$Kansas State University, Manhattan, KS 66506,
$^p$University of Manchester, Manchester M13 9PL, England,
$^q$Queen Mary, University of London, London, E1 4NS, England,
$^r$Muons, Inc., Batavia, IL 60510,
$^s$Nagasaki Institute of Applied Science, Nagasaki, Japan, 
$^t$National Research Nuclear University, Moscow, Russia,
$^u$University of Notre Dame, Notre Dame, IN 46556,
$^v$Universidad de Oviedo, E-33007 Oviedo, Spain, 
$^w$Texas Tech University, Lubbock, TX  79609, 
$^x$IFIC(CSIC-Universitat de Valencia), 56071 Valencia, Spain,
$^y$Universidad Tecnica Federico Santa Maria, 110v Valparaiso, Chile,
$^z$University of Virginia, Charlottesville, VA  22906,
$^{aa}$Yarmouk University, Irbid 211-63, Jordan,
$^{ii}$On leave from J.~Stefan Institute, Ljubljana, Slovenia, 
}}
\noaffiliation
\vspace*{4em}

%% file: WZprd_final_eprint.bbl
\newpage
\begin{thebibliography}{99} 
\bibitem{Collins}P.~D.~B. Collins, {\em An Introduction to Regge Theory and High Energy Physics}, Cambridge University  Press, Cambridge, (1977).
\bibitem{Barone}V.~Barone and E.~Predazzi, {\em High-Energy Particle Diffraction},
Springer Press, Berlin, (2002).
\bibitem{Donnachie}
S.~Donnachie, G.~Dosch, P.~Landshoff, and O.~Nachtmann, {\em Pomeron Physics and QCD}, Cambridge University Press, Cambridge, (2002). 
\bibitem{rapidity}
Rapidity, $y=\frac{1}{2}\ln\frac{E+p_L}{E-p_L}$, and pseudorapidity, $\eta=-\ln\tan\frac{\theta}{2}$, where $\theta$ is the polar angle of a particle with respect to the proton beam ($+\hat z$ direction), are used interchangeably for particles detected in the calorimeters, 
since in the kinematic range of interest in this analysis they are approximately equal.
\bibitem{run1dijet1997}F. Abe {\em et al.} (CDF Collaboration), Phys. Rev. Lett. {\bf 79}, 2636 (1997).
 \bibitem{ken_thesis}K.~Hatakeyama, Ph.D. thesis, The Rockefeller University, 2003; FERMILAB-THESIS-2003-18.
\bibitem{run1dijet2000}T. Affolder {\em et al.} (CDF Collaboration), Phys. Rev. Lett {\bf 84}, 5043 (2000).
\bibitem{run1dijet2002}D. Acosta {\em et al.} (CDF Collaboration), Phys. Rev. Lett. {\bf 88}, 151802 (2002). 
\bibitem{run1}F. Abe {\em et al.} (CDF Collaboration), Phys. Rev. Lett. {\bf 78}, 2698 (1997).
\bibitem{cdfdiffrb}T. Affolder {\em et al.} (CDF Collaboration), Phys. Rev. Lett. {\bf 84}, 232 (2000). 
\bibitem{jpsi}T. Affolder {\em et al.} (CDF Collaboration), Phys. Rev. Lett. {\bf 87}, 241802 (2001).
\bibitem{d0} V. M. Abazov {\em et al.} (D0 Collaboration), Phys. Lett. B {\bf 574}, 169 (2003).
\bibitem{AlanWhite}A.~R.~White, Phys. Rev. {\bf D72}, 036007 (2005).
\bibitem{cdf}A.~Abulencia {\it et al.} (CDF Collaboration), J.\ Phys.\ G {\bf 34}, 2457 (2007).
\bibitem{DPEdijet}T. Aaltonen {\em et al.} (CDF Collaboration), Phys. Rev. D {\bf 77}, 052004 (2008).
\bibitem{Wcrosssect}D. Acosta {\em et al.} (CDF Collaboration), Phys. Rev. Lett. {\bf 94}, 091803 (2005).
\bibitem{missingET} Transverse energy is defined as $E_T=E\sin\theta$, and missing $E_T$ as $\met = |\overrightarrow{\met}|$ with  $\overrightarrow{\met}= -\sum_iE_T^i\hat{n_i}$, where $\hat{n_i}$ is a unit vector perpendicular to the beam axis and pointing at the $i^{th}$ calorimeter tower. The sum $E_T$ is defined by $\sum E_T=\sum_iE_T^i$. Both sums are over all calorimeter towers above the set thresholds.\bibitem{CDF_elastic}F.~Abe {\em et al.} (CDF Collaboration), Phys. Rev. D {\bf 50}, 5518 (1994).
\bibitem{CDF_total}F.~Abe {\em et al.} (CDF Collaboration), Phys. Rev. D {\bf 50}, 5550 (1994).
\bibitem{PDG}C. Amsler {\em et al.} (Particle Data Group), Phys. Lett. B {\bf 667}, 1 (2008) and 2009 partial update for the 2010 edition.
\bibitem{Barlow} R. Barlow and C. Beeston, Comput. Phys. Commun. {\bf 77}, 219 (1993).
\bibitem{exclZ}T. Aaltonen {\em et al.} (CDF Collaboration), Phys. Rev. Lett. {\bf 102}, 222002 (2009).
\bibitem{albrow_forshaw}M.~G.~Albrow, T.~D.~Coughlin and J.~R.~Forshaw, arXiv:1006.1289 [hep-ph], Prog. Part. Nucl. Phys. (to be published).
\end{thebibliography}
